# A two-stage prediction model for heterogeneous effects of treatments

Konstantina Chalkou[1], Ewout Steyerberg[2], Matthias Egger[1,3], Andrea Manca[4], Fabio Pellegrini[5], Georgia Salanti[1]

**Affiliations:** Institute of Social and Preventive Medicine, University of Bern, Bern, Switzerland [1]; Leiden University Medical Center, Leiden, the Netherlands [2]; Population Health Sciences, Bristol Medical School, University of Bristol, Bristol, UK [3]; Centre for Health Economics, University of York, York, UK [4]; Biogen International GmbH, Baar, Switzerland [5]

Correspondence to:
Konstantina Chalkou, MSc
Institute of Social & Preventive Medicine
University of Bern
Mittelstrasse 43, 3012 Bern, Switzerland
konstantina.chalkou@ispm.unibe.ch

*Abstract*

Treatment effects vary across different patients, and estimation of this variability is essential for clinical decision-making. We aimed to develop a model estimating the benefit of alternative treatment options for individual patients, extending a risk modelling approach in a network meta-analysis framework. We propose a two-stage prediction model for heterogeneous treatment effects (HTE) by combining prognosis research and network meta-analysis methods where individual patient data are available. In the first stage, a prognostic model to predict the baseline risk of the outcome. In the second stage, we use the baseline risk score from the first stage as a single prognostic factor and effect modifier in a network meta-regression model. We apply the approach to a network meta-analysis of three randomized clinical trials comparing the relapses in Natalizumab, Glatiramer Acetate and Dimethyl Fumarate, including 3590 patients diagnosed with relapsing-remitting multiple sclerosis. We find that the baseline risk score modifies the relative and absolute treatment effects. Several patient characteristics, such as age and disability status, impact the baseline risk of relapse, which in turn moderates the benefit expected for each of the treatments. For high-risk patients, the treatment that minimises the risk of relapse in two years is Natalizumab, whereas Dimethyl Fumarate might be a better option for low-risk patients. Our approach can be easily extended to all outcomes of interest and has the potential to inform a personalised treatment approach.

## 1 Introduction

Personalised predictions are important for clinical decision-making. The question 'Which treatment is best?' can have two very different meanings: 'Which treatment is best on average?' or 'Which treatment is best for a specific patient?' Patients often experience different outcomes under the same treatment. One patient may benefit more from a treatment from which another patient may benefit less. Thus, it is essential to identify the patient characteristics that influence treatment effects to choose a given patient's best option. Prediction models aim to identify and estimate the impact of patient, intervention and setting characteristics on future health outcomes.

Effect modification and risk modelling approaches estimating heterogeneous treatment effects are gaining ground in meta-analysis.[1,2,3] Effect modification predicts individualised treatment effects via a model that incorporates a term for treatment assignment and treatment by covariate interaction terms.[1,2,4] However, selecting effect modifiers in a meta-analysis context is challenging for many reasons. These include low power and overfitting, misleading estimates because of unreliable, exaggerated, and highly influential interaction terms and the risk of discovering false subgroup effects because of weak prior knowledge.[1,5,6,7] Also, guidance is missing about the model selection techniques and shrinkage methods in meta-regression models optimal to examine effect modification. Alternatively, modellers can take advantage of the fact that patients' baseline risk is often a determinant of heterogeneous treatment effects.[1,7,9,10] A risk modelling approach predicts the risk for patients based on their baseline characteristics. It then uses this risk to predict heterogeneous treatment effects at the absolute scale, typically within a randomised clinical trial (RCT).[1,2,7,8,11,12,13] In this sense, risk modelling deals better with dimensionality, low power and limited prior knowledge than an effect modification approach. However, its use constrains the model's flexibility, as all prognostic factors also act as effect modifiers via a single coefficient.[2]

The baseline risk expresses the probability of experiencing the outcome of interest in the study. Models that link the baseline risk to patient characteristics have been referred to as *prognostic* or *risk models*. These models can be integrated with the risk modelling approach. The first step is to develop a multivariable prognostic model that predicts the probability of the studied outcome blinded to the treatment - this can be done using observational or RCT data. We will term this *baseline risk* from now on, and a transformation of this risk will be termed *baseline risk score*. Several established methods exist for developing a prognostic model.[14,15,16,17] In the second step, relative treatment effects within RCTs can be estimated as a function of the baseline risk score using a prediction model.[18] This methodology allows for heterogeneity in baseline risk, in the relative treatment effects and consequently in the absolute treatment effects. The risk modelling approach has recently gained ground for personalised predictions for a given treatment.[1,11]

Multiple sclerosis (MS) is an autoimmune disease of the central nervous system with several subtypes. The most common subtype is relapsing-remitting Multiple Sclerosis (RRMS).[19] Patients with RRMS present with intense symptoms (relapses) followed by periods without symptoms (remission).[20] Several treatments are available, but patient responses are heterogeneous, and each treatment has a different safety profile.[21]

The evidence on drugs for RRMS has been summarised using network meta-analysis.[22,23] These networks typically synthesise published aggregated data, and their ability to explore how patient characteristics influence treatment effects (relative or absolute) across different patients is limited.[24] More efficient analyses use individual patient data (IPD), considered the gold standard in evidence synthesis.[24] IPD are necessary for estimating heterogeneous treatment effects and making personalised predictions of expected outcomes.[24,25]

This paper aims to define a methodological framework that allows personalised predictions for the most likely outcome under several treatment options. To achieve this, we adapt the risk modelling approach for the context of meta-analysis, extending it to a network meta-analysis framework. We combine prognostic modelling ideas to estimate the baseline risk score and include this score in an IPD network meta-regression (NMR). We apply this method to a set of placebo-controlled trials of three drugs in patients with RRMS. We also examine how different prognostic models to estimate the baseline risk score influence the predictive model's results and the estimated absolute and relative treatment effects.[15,26] We present results primarily for the absolute treatment effects. These will vary across patient groups, even if heterogeneity is present only in the baseline risk but not in the relative treatment effects. We describe the general framework applicable to any type of data and network, along with the detailed methods for our application to drugs for RRMS.

## 2 Methods

In this section, we present a general description of the two-stage model, where we first obtain the baseline risk score and then estimate outcomes' probabilities as a function of the score. The baseline risk score is determined using established methods for the predictors' selection (e.g. pre-specified, stepwise, and penalised methods), for the estimation and shrinkage of the coefficients (e.g. uniform, elastic net, and penalised maximum likelihood estimation method), and for its validation and presentation.[15,26,27] In the second stage, we used an IPD NMR model with the baseline risk score, developed in stage 1, as prognostic factor and effect modifier of the outcome. Our approach assumes that the set of selected variables captures both prognosis and effect modification adequately. We describe the approach for a dichotomous outcome of interest, although continuous outcomes can also be modelled with minor modifications.

Along with the general description of the framework, we describe an application of our methodology, which predicts relapses in two years for individuals diagnosed with RRMS. In section 2.1, we describe the data we used, and in 2.2, we present the notation used in our statistical models. In sections 2.3 and 2.4, we present the first and second stages and the methods implemented in the example. Finally, we present in section 2.5 the software and the functions used. In our application, we chose to implement the first stage in a frequentist framework to take advantage of the shrinkage options readily available in software and the second stage in a Bayesian framework.

## *2.1*     Data description

We analysed IPD from three phase III randomised clinical trials: AFFIRM,[28] DEFINE,[29] and CONFIRM [30] on patients diagnosed with RRMS. Altogether, the trials included 3590 patients randomised to placebo, Natalizumab, Dimethyl Fumarate, and Glatiramer Acetate. The outcome of interest was relapse or not relapse in two years. Table 1 presents the aggregated-level data of the trial arms as well as some baseline characteristics. We also had access to IPD from 1083 patients with RRMS, randomised to placebo arms included in nine other clinical trials. The latter data were provided by the Clinical Path Institute (https://c-path.org/) and are also described in Table 1. We excluded variables with more than 50% missing values. We used complete case analysis for the remaining variables, assuming that any missingness does not depend on the risk of relapse. We think this is reasonable as all variables are measured at baseline, and the outcome is observed in a two-year's time window. Between correlated variables (correlation coefficient larger than 70%), we retained those that were biologically plausibly associated with the outcome based on the literature, their distribution and the amount of missing values. Finally, we transformed some of the continuous variables to approximate the normal distribution and merged categories with very low frequencies in categorical variables.

Table 1. Baseline characteristics of relapsing-remitting multiple sclerosis (RRMS) patients enrolled in the trials

| Study | Treatment | Number of randomized patients | Number of patients with relapse in two years | Age | Sex | | Baseline EDSS | Number of relapses in previous year |
|---|---|---|---|---|---|---|---|---|
| | | | | Mean (sd) | Female N (%) | Male N (%) | Mean (SD) | Median (min, max) |
| **AFFIRM** | | 939 | 359 (38.2%) | 36.0 (8.3) | 657 (70.0) | 282 (30.0) | 2.3 (1.2) | 1 (0, 12) |
| | Natalizumab | 627 | 183 (29.2%) | | | | | |
| | Placebo | 312 | 176 (56.4%) | | | | | |
| **CONFIRM** | | 1417 | 451 (31.8%) | 37.3 (9.3) | 993 (70.1) | 424 (29.9) | 2.6 (1.2) | 1 (0, 8) |
| | Dimethyl Fumarate | 703 | 185 (26.3%) | | | | | |
| | Glatiramer Acetate | 351 | 117 (33.3%) | | | | | |
| | Placebo | 363 | 149 (41.0%) | | | | | |
| **DEFINE** | | 1234 | 394 (31.9%) | 38.5 (9.0) | 908 (73.6) | 326 (26.4) | 2.4 (1.2) | 1 (0, 6) |
| | Dimethyl Fumarate | 826 | 212 (25.7%) | | | | | |
| | Placebo | 408 | 182 (44.6%) | | | | | |
| **Placebo arms dataset** | Placebo | 1083 | 801 (74.0%) | 41.19 (10.3) | 752 (69.4) | 331 (30.6) | NA | NA |

EDSS: Expanded Disability Status Scale; NA: Not available

### 2.2 Notation

Let $Y_{ij}$ denote the dichotomous outcome for individual $i$ where $i=1, 2, ..., n_j$ in the $j$ study out of $ns$ trials. $PF_{ijk}$ is the $k$ prognostic factor and $np$ is the total number of prognostic factors. An individual can develop the outcome ($Y_{ij} = 1$) or not ($Y_{ij} = 0$) according to their risk at baseline, which is a function of the prognostic factors, and we denote it with $R_{ij}$. Assume we have a set of treatments $\mathcal{H}$ each denoted by $t \in \mathcal{H}$ where $t = 1, 2, \ldots, T$. The probability $p_{ijt}$ is the probability of the outcome for the $i$ individual in $j$ study under treatment $t$ and depends on the treatment, baseline risk score and the interaction between the risk score and the treatment.

### 2.3 Stage 1: Developing a baseline risk score model

We developed risk models for dichotomous outcomes using two different methods. The first model was selected via the LASSO (Least Absolute Shrinkage and Selection Operator) method. The second used a pre-specified risk model.[26] Observational or RCT data may be used for this purpose. For the application, only placebo-controlled RCTs were available. Following the PATH recommendation,[32] when developing a baseline risk score using RCTs, not only the placebo arms but the entire trial population blinded to the treatment should be used.[7,31,32] Using only placebo arms only decreases the effective sample size. It may also lead to differential model fit on trial arms, biasing treatment effect estimates across risk strata, and exaggerating HTE.[1,7,32,33] It can make completely ineffective treatments appear to be beneficial in high-risk patients and harmful in low-risk patients.[2]

The logistic regression model is

$$Y_{ij} \sim Bernoulli(R_{ij})$$

$$logit(R_{ij}) = b_{0j} + \sum_{j=1}^{np} b_{kj} \times PF_{ijk} \quad (1)$$

The regression coefficients and intercept can be independent (each $b_{0j}$ is given a prior distribution), exchangeable ($b_{0j} \sim N(\beta_0, \sigma_{B_0}^2)$, $b_{kj} \sim N(\beta_k, \sigma_{B_k}^2)$) or common ($b_{0j} = \beta_0$, $b_{kj} = \beta_k$) across studies. For model selection, methods that include some form of penalisation are preferred to stepwise selection.[14,15,26] The latter include LASSO. However, including a set of predictors informed by prior knowledge (either in the form of expert opinion or previously identified variables in prognostic studies) has conceptual and computational advantages.[26,27,34] The estimated effects of the selected covariates also need some form of penalisation to avoid extreme predictions.[14,15] In the illustration of our empirical example, we discuss several possibilities.

In our empirical example, we developed a baseline risk model for relapse in two years. We first examined if the available sample size was enough for the development of a prognostic model.[35] We calculated the events per variable (EPV), accounting for categorical variables and non-linear continuous variables.[36] We also used Riley at al.'s method to calculate the required minimum sample size for a logistic model.[37] We set Nagelkerke's $R^2$ = 0.15 (Cox-Snell's adjusted $R^2$ = 0.11) and the desired shrinkage equal to 0.9.

We then fitted two main prognostic models. In the first, we included predictors with non-zero coefficients in the LASSO.[38] We used the LASSO method both for the variable selection and for estimating the coefficients. We used 10-fold cross-validation to find the optimal penalty parameter that maximises the area under the curve. The penalty parameter we chose is the one within one standard error of the minimum parameter, as previously recommended.[15]

The second prognostic model included previously identified prognostic factors. Pellegrini et al. analysed the annualised relapse rate in the DEFINE (training dataset) and CONFIRM (validation dataset) trials,[39] both of them included in our dataset as described in section 2.1. They used different modelling approaches, including a fully additive model, ridge regression, LASSO, and elastic net regression. They selected the additive model, including 14 prognostic factors based on its discriminative ability. We estimated the coefficients in each of these prognostic factors in our dataset (section 2.1), using penalised maximum likelihood estimation shrinkage method.[15,40] The penalty's optimal value was chosen as the one that maximises a modified Akaike's Information Criterion.[15] Both models use common effects for the intercept and the regression coefficients ($b_{0j} = \beta_0$, $b_{kj} = \beta_k$). This decision was taken because all three trials were designed by the same company using a similar protocol, as described in section 2.1, and any differences in the included populations shall be captured by including the baseline risk in the NMR model.

Finally, validation is essential for evaluating the performance of a prognostic model.[16] As external data were not available, we performed internal validation only. We estimated the c-statistic and the calibration slope of the developed risk models to assess the discriminative performance and calibration. To account for optimism, which is particularly important when comparing various models, we used the bootstrap method.[15] We produced 500 bootstraps samples and reran the model selection process and estimation in each sample. Then, we assessed the performance of each bootstrap-based model in the original sample.[41,42]

## 2.4   Stage 2: IPD Network meta-regression model

We used the baseline risk logit as a covariate in an IPD NMR model in the second stage.[43] Each study $j$ has an arbitrarily chosen baseline treatment $h_j \in \mathcal{H}$ and then each individual $i$ is randomised to any treatment $t \in \mathcal{H}$ included in study $j$. The meta-regression equation in study $j$ with a baseline treatment $h_j$ will be:

$$Y_{ij} \sim Bernoulli(p_{ijt})$$

$$logit(p_{ijt}) = \begin{cases} u_j + g_{0j} \times \left(logit(R_{ij}) - \overline{logit(R_{ij})}^j\right) & \text{if } t = h_j \\ u_j + d_{jh_jt} + g_{0j} \times \left(logit(R_{ij}) - \overline{logit(R_{ij})}^j\right) + g_{jh_jt} \times \left(logit(R_{ij}) - \overline{logit(R_{ij})}^j\right), & \text{if } t \neq h_j \end{cases}$$

(2)

where $\overline{logit(R_{ij})}^j$ is the average of logit-risk in all individuals in study $j$ and $u_j$ is the log-odds for the reference treatment arm when the logit-risk is equal to $\overline{logit(R_{ij})}^j$. The nuisance parameters $u_j$ are then considered to be independent. The relative treatments effects are the log-odds ratios $d_{jh_jt}$ and can be random ($d_{jh_jt} \sim N(D_{h_jt}, \sigma_D^2)$) or common ($d_{jh_jt} = D_{h_jt}$) across studies. Then, assuming consistency, we set the constraint $D_{h_jt} = \delta_t - \delta_{h_j}$ and $\delta_{ref} = 0$ where $\delta_t$ is the summary estimate for log-odds ratios for treatment t versus the overall reference treatment (denoted as $ref$). Parameter $g_{0j}$ is the coefficient of the risk score (as a prognostic factor) and should be independent across studies (so that each $g_{0j}$ is given a prior distribution) to avoid compromising randomisation. In case of model non-convergence or studies following the same protocol (as in our example), exchangeable ($g_{0j} \sim N(\gamma_0, \sigma_{\gamma_0}^2)$), or common ($g_{0j} = \gamma_0$) coefficients can be used.[44,45] Similarly, $g_{jh_jt}$ refers to the treatment effect modification of the risk score, for treatment $t$ versus study's baseline treatment $h_j$, and can be random ($g_{jh_jt} \sim N(G_{h_jt}, \sigma_G^2)$) or common ($g_{jh_jt} = G_{h_jt}$). Similarly to the relative treatment effects, the regression coefficients $G_{h_jt}$ between two active treatments are parametrized using basic parameters $\gamma_t$ (of each active treatment versus control), where $G_{h_jt} = \gamma_t - \gamma_{h_j}$ and $\gamma_{ref} = 0$. Finally, $\exp(\gamma_t)$ is the ratio of two odds ratios (ORs) of treatment $t$ versus the reference: the OR of a group of people with baseline score $x$ over the OR in a group of people with baseline risk score $x - 1$.

Assume an overall reference treatment (like placebo or the current standard treatment) for which predictions are less important. Then, consider a patient at the mean (logit) baseline population risk, $\bar{R}$ who is under the reference treatment. This logit-probability of the outcome is denoted with, say $a$. To make predictions for a new patient with predicted risk $\widetilde{logit(R_\iota)}$ and in treatment $t$, we use the equation:

$$logit(p_i) = a + \delta_t + \gamma_0 \times \left(\widetilde{logit(R_i)} - \overline{logit(R)}\right) + \gamma_t \times \left(\widetilde{logit(R_i)} - \overline{logit(R)}\right), \quad (3)$$

Estimation of $a$ and $\overline{logit(R)}$ depends on the context within which we plan to make predictions: one can use registry data, observational studies or RCT data. For example, $\overline{logit(R)}$ can be estimated as the mean of $logit(R_{ij})$ across all individuals in the (randomized or observational) studies. Similarly, $a$ can be estimated from the synthesis of all untreated or placebo arms.

In our empirical example, we used an IPD NMR model for comparing three active treatments and placebo in RRMS patients, using the predicted risk obtained from the first stage (LASSO and pre-specified model), $logit(R_i)$. We assume that study-specific relative treatment effects do not have any residual heterogeneity beyond what is already captured by differences in baseline risk. Consequently, we employ a common effect IPD NMR model, both in the relative treatment effects $d_{jb_jt}$ and for the treatment effect modification of the risk score. Note that the between studies variance could not be estimated with only three studies ($d_{jh_jt} = D_{h_jt} = \delta_t - \delta_{h_j}$, $\delta_{ref} = 0$, $g_{jh_jt} = G_{h_jt} = \gamma_t - \gamma_{h_j}$, $\gamma_{ref} = 0$). We also assumed common coefficients for the risk score ($g_{0j} = \gamma_0$), as all three studies are very similar in terms of design characteristics.

To estimate the logit-probability ($a$) of the outcome of a patient in placebo who has a baseline risk score equal to the average risk $\overline{logit(R)}$ we synthesised external IPD placebo-arm data.

## 2.5 Implementation and software

All our analyses were done in R,[46] using R 3.6.2 version and in JAGS [47] (called through R). We make the code available in a GitHub library: https://github.com/htx-r/Reproduce-results-from-papers/tree/master/ATwoStagePredictionModelMultipleSclerosis.

To develop the baseline risk model (2.3), we used the `pmsampsize` command to estimate if the available sample size was enough for the developed model. The LASSO model was developed using `cv.glmnet`. We first fitted the pre-specified model using the

`lrm` command and used the `pentrace` command for the penalised maximum likelihood estimation. For the bootstrap internal validation, we used self-programmed R-routines.

The IPD NMR model (2.4) was fitted in a Bayesian framework, and we used programming routines in the `R2Jags` package.[48] We set a normal distribution ($N(0,1000)$) as prior distributions for all of the model parameters. We simulated two chains of 10,000 samples, discarded the first 1,000 samples and thinned for every 10 samples. This was deemed appropriate based on autocorrelation plots and the visualisation of the chain convergence.

## 3 Results

### 3.1 Stage 1: Developing the baseline risk score

A total of 57 candidate prognostic factors were available. After exclusion of variables with missing data and highly correlated data, we ended up with 31 candidate prognostic factors (Appendix figure 1, Appendix figure 2).

For the LASSO model, we used 2000 RRMS patients with complete data, 742 of whom relapsed in two years. The full model had 45 degrees of freedom, and the EPV was 16.5. The recommended sample size for a newly developed model is 3456 patients, which is more than the available sample size. For the pre-specified model, which does not involve the selection of variables, the small number of degrees of freedom (14) led to a large EPV of 53 and a recommended minimum sample size of 1076, which is well below the available sample size.

Table 2 shows the two models, their coefficients and their performance with internal validation. Both models have almost the same discriminative ability, but the pre-specified model has a much better calibration slope.

Both models predict almost the same mean risk for patients in our data (about 37%), as shown in Figure 1. The variation in the estimated baseline risk score is much higher in the pre-specified model, using the predictors of Pellegrini et al.[39] Figure 1 also indicates that the baseline risk could be a prognostic factor for relapse, as the baseline risk score is higher for patients who relapsed than for patients who did not, using both models. However, the overlap is considerable, as also shown by the c-statistics in Table 2.

**Table 2. Estimated LASSO (Least Absolute Shrinkage and Selection Operator) shrunk coefficients and coefficients from the pre-specified model together with penalized maximum likelihood estimation. The discrimination (C-score) and the calibration slopes are also shown.**

| Variables | LASSO model Coefficients | Pre-specified model Coefficients (S.E.) |
|---|---|---|
| **C-score** | 0.60 | 0.62 |
| **Calibration slope** | 1.54 | 1.05 |
| Intercept | -0.4424 | -0.8656 (0.866) |
| Age | -0.0013 | -0.0181 (0.005) |
| Sex (male vs female) | - | -0.1379 (0.092) |
| Baseline weight | -0.0002 | - |
| Baseline EDSS | 0.0963 | 0.1683 (0.047) |
| Years Since Onset of Symptoms | - | 0.0587 (0.063) |
| Ethnicity (white vs other) | - | -0.0142 (0.117) |
| No. of relapses 1 year prior to study | 0.2971 | 0.5963 (0.170) |
| Months since pre-study relapse | - | -0.0126 (0.009) |
| Prior MS treatment group (yes vs no) | 0.0241 | 0.1901 (0.085) |
| Region (India vs Eastern Europe) | 0.0000 | - |
| Region (North America vs Eastern Europe) | 0.0000 | - |
| Region (Rest of world vs Eastern Europe) | 0.0000 | - |
| Region (Western Europe vs Eastern Europe) | 0.2374 | - |
| Timed 25-Foot Walk | - | -0.1718 (0.158) |
| 9-Hole Peg Test | - | 0.3011 (0.208) |
| PASAT-3 | - | 0.0029 (0.004) |
| VFT 2.5% | - | -0.0010 (0.004) |
| Baseline Gadolinium volume | 0.0001 | - |
| Baseline SF-36 PCS | -0.0120 | -0.0195 (0.005) |
| Baseline SF-36 MCS | - | 0.036 (0.004) |
| Baseline Actual Distance Walked (>500 vs <=500) | -0.0746 | - |

S.E: Standard Error; EDSS: Expanded Disability Status Scale; MS: multiple sclerosis; PASAT: Paced Auditory Serial Addition Test; VFT: Visual Function Test; SF-36 PCS: Short Form-36 Physical Component Summary; SF-36 MCS: Short Form-36 Mental Component Summary

**Figure 1** The distribution of the baseline risk for LASSO model (A) and pre-specified model (B) for patients that did not relapse in two years and for patients that did relapse in two years. The dotted lines indicate group means and the solid line the overall mean risk

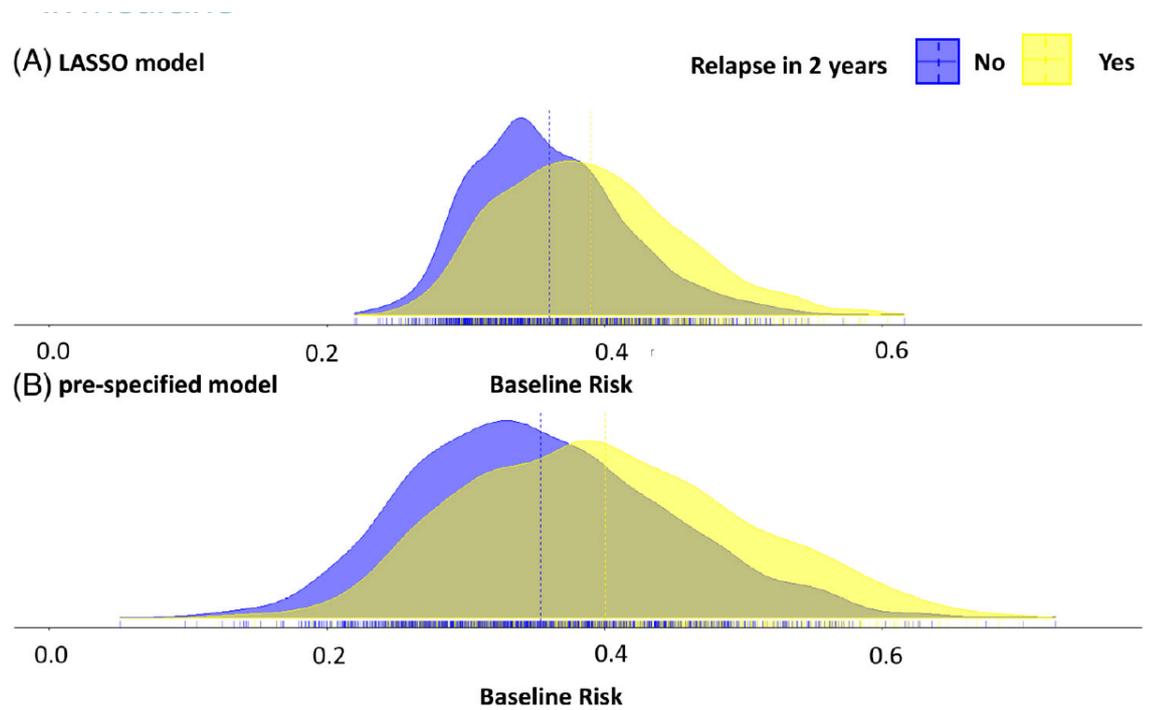

### *3.2* Stage 2: Estimating heterogeneous treatment effects in an IPD network meta-regression model

Table 3 shows the estimated parameters from the network meta-regression model using the two different scored developed from the LASSO model and pre-specified model. Both models indicate the baseline risk as an important prognostic factor for relapsing at two years, as shown by the large values for $\gamma_0$. The estimates of log ORs for each treatment versus placebo ($\delta_t$) are very similar with both models. However, they provide slightly different summary estimates for the coefficients of effect modification, i.e. $\gamma_{DF}, \gamma_{GA}, \gamma_N$. Overall, none of the coefficients $\gamma_{DF}, \gamma_{GA}, \gamma_N$ is large.

Figure 2 shows the estimated predicted probabilities to relapse within two years depending on the estimated baseline risk, via LASSO and pre-specified risk models, under the four available treatment options. Appendix figure 3 presents the same results on the OR scale. Both models give almost the same results for the treatment-effects estimation: Glatiramer Acetate seems to have the same performance as Dimethyl Fumarate in the observed range of baseline risk; placebo results in the highest risk to relapse. Natalizumab is a drug initially considered less safe than the other two active options and associated with increased mortality.[49,50] Table 4 shows the estimated predicted probabilities and the ORs of relapsing under all three available active treatments, using both models separately, for all patients, for low-risk patients (baseline risk <30%) and for high-risk patients (baseline risk >50%). The benefit of all three treatments depends on the risk group. For high-risk patients, the absolute benefit of Natalizumab compared to Dimethyl Fumarate is 15% using the pre-specified model and 10% for the LASSO model. These correspond into 7 and 10 patients respectively that need to be treated with Natalizumab to prevent one relapse. For low risk patients, the absolute benefit of Dimethyl Fumarate compared to Natalizumab is 3% for the pre-specified model and 2% for the LASSO model. The absolute differences between the treatments for all risk-groups are smaller using LASSO compared to the (penalised) pre-specified model. The predictions for the three drugs and placebo for RRMS have been implemented in an interactive R-Shiny application available at https://cinema.ispm.unibe.ch/shinies/koms/.

**Table 3. Estimated parameters from the network meta-regression model using the two different scores developed from the LASSO model and pre-specified model**

| Estimated parameters from IPD NMR model | LASSO model Mean (95% Cr. Interval) | Pre-specified model Mean (95% Cr. Interval) |
|---|---|---|
| $\gamma_0$ | 2.30 (1.78, 2.8) | 1.26 (0.95, 1.58) |
| $\delta_{DF}$ | -0.92 (-1.20, -0.64) | -0.89 (-1.18, -0.60) |
| $\delta_{GA}$ | -0.72 (-1.15, -0.28) | -0.71 (-1.15, -0.26) |
| $\delta_N$ | -1.24 (-1.55, -0.93) | -1.22 (-1.53, -0.93) |
| $\gamma_{DF}$ | 0.90 (-0.20, 1.98) | 0.25 (-0.35, 0.87) |
| $\gamma_{GA}$ | 0.64 (-1.02, 2.39) | 0.23 (-0.71, 1.3) |
| $\gamma_N$ | -0.02 (-1.16, 1.07) | -0.26 (-1.01, 0.43) |

$e^{\gamma_0}$: OR of relapse in two years for one unit increase in logit-risk in untreated patients (placebo)
$e^{\delta_{DF}}$: OR of relapse under Dimethyl Fumarate versus placebo at the study mean risk
$e^{\delta_{GA}}$: OR of relapse under Glatiramer Acetate versus placebo at the study mean risk
$e^{\delta_N}$: OR of relapse under Natalizumab versus placebo at the study mean risk
$e^{\gamma_{DF}}$: OR of relapse under Dimethyl Fumarate versus placebo for one unit of increase in the logit risk
$e^{\gamma_{GA}}$: OR of relapse under Glatiramer Acetate versus placebo for one unit of increase in the logit risk
$e^{\gamma_N}$: OR of relapse under Natalizumab versus placebo for one unit of increase in the logit risk
DF: Dimethyl fumarate; GA: Glatiramer acetate; N: Natalizumab

**Figure 2** The distribution of the baseline risk for LASSO model (A) and pre-specified model (B) for patients that did not relapse in two years and for patients that did relapse in two years. The dotted lines indicate group means and the solid line the overall mean risk

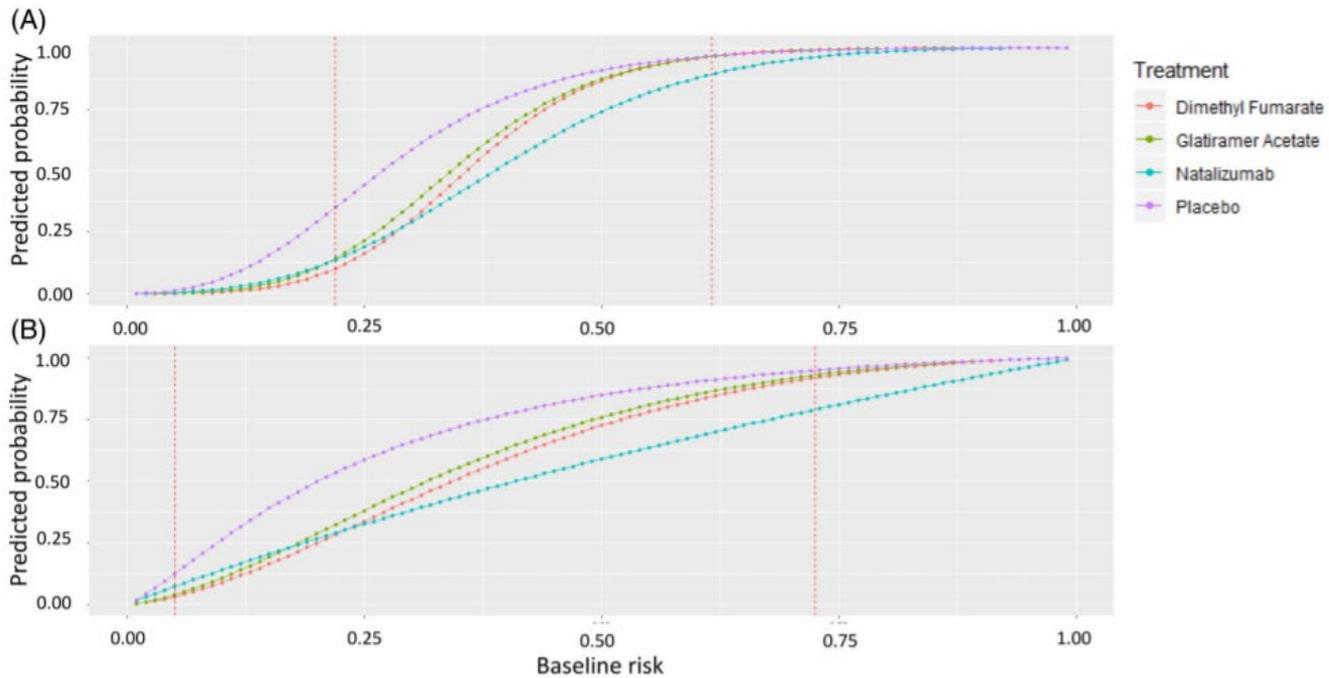

**Table 4.** Predicted % probabilities and odds ratios (ORs, relative benefits) of relapse in two years, using baseline risk scores developed with the LASSO (Least Absolute Shrinkage and Selection Operator) and pre-specified models. Results are shown for all patients, for low-risk patients (baseline risk <30%) and for high-risk patients (baseline risk >50%) in the observed range of baseline risk. The cut-offs have been chosen arbitrarily for illustrative purposes.

| Benefits | Model | Treatment | All patients | Baseline risk <30% Low-risk patients | Baseline risk >50% High-risk patients |
|---|---|---|---|---|---|
| Absolute benefits (%) | LASSO | Dimethyl Fumarate | 62% | 18% | 93% |
| | | Glatiramer Acetate | 64% | 23% | 93% |
| | | Natalizumab | 54% | 20% | 82% |
| | Prespecified | Dimethyl Fumarate | 53% | 20% | 84% |
| | | Glatiramer Acetate | 56% | 23% | 86% |
| | | Natalizumab | 46% | 23% | 69% |
| Relative benefits (OR) | LASSO | Dimethyl Fumarate vs placebo | 0.52 | 0.25 | 0.81 |
| | | Glatiramer Acetate vs placebo | 0.57 | 0.35 | 0.81 |
| | | Natalizumab vs placebo | 0.29 | 0.29 | 0.28 |
| | Prespecified | Dimethyl Fumarate vs placebo | 0.42 | 0.31 | 0.53 |
| | | Glatiramer Acetate vs placebo | 0.50 | 0.38 | 0.63 |
| | | Natalizumab vs placebo | 0.31 | 0.40 | 0.23 |

## 4  Discussion

We developed a prediction model for heterogeneous treatment effects that combines risk modelling and network meta-analytical methods to make personalised predictions for an outcome of interest. As the treatment options for each condition are numerous and patient characteristics often play an important role in modifying treatment effects, this approach could contribute to personalized treatment decisions. We illustrated our method by comparing three active treatments and placebo in patients with RRMS. Only a few characteristics are required, and doctors and patients can enter these into our online tool (https://cinema.ispm.unibe.ch/shinies/koms/) to estimate the risk of relapse in two years under four treatment options.

The application to RRMS shows the approach's potential but is not ready for use in clinical practice. Decision-making tools need external validation with new patients. They need to provide evidence about all available treatment options for patient-relevant outcomes (e.g. long-term disability status[51]) and also consider safety and costs. Unfortunately, long-term results are not available from RCTs, and observational data would need to be integrated for this purpose. We did not have access to such data, which would have also allowed us to validate the model externally. Because of the limited data availability (only three RCTs), we used common-effects IPD-NMR to facilitate model convergence. The common-effects assumption can be relaxed if more studies are available. Making personalised predictions using a random-effects model will increase the uncertainty, and the interpretation of results in the presence of large heterogeneity will be challenging. Another extension of our model could accommodate aggregate-level data to increase the relevant information and ensure that the findings are representative. This is particularly important when the analyst is interested in comparing all available treatments and making corresponding predictions.[43]

We aimed to examine whether and by how much the risk modelling results are influenced by the method used to develop the baseline risk model (i.e. stage 1). A pre-specified model used variables previously identified as important prognostic factors.[39] Additionally, we used a variable selection approach via LASSO. The two models in stage 1 differ in terms of included variables, however the choice of the model had only a small impact on the results of stage 2. Whether this is a general feature of the approach or this happen to be true in this particular dataset should be subject of further research. More applications and a simulation study would be needed to pinpoint the sensitivity of the final results to the choice of the model in the first stage. The models' discrimination was small but sufficient for our aim. Indeed, risk models with a low predictive ability (0.6 – 0.65) are often adequate to detect risk-based heterogeneous treatment effects.[11] The available sample size was sufficient for the pre-specified model (as it did not involve variable selection), whereas it was not for the LASSO model. The pre-specified model's discriminative ability was slightly better, and the calibration slope much better than the LASSO model. This finding corroborates previous guidance in the literature that suggests that use of prior evidence in model development is advantageous.[2,26]

The approach has several limitations. Our framework requires at least one IPD dataset for each included intervention to estimate all model parameters. IPD data are not readily available: several papers have documented the difficulties encountered in the process.[52,53,54] When condensing all patient information into the risk score, we assume that the selected variables adequately capture both prognosis and effect modification. This assumption is difficult to evaluate unless the outcome is well studied, and many prognostic studies exist on the topic, which is rarely the case. Besides, it is possible that other study-level characteristics, such as the risk of bias and the year of randomisation, may also impact the treatment effects. If the number of studies permits, such variables can be added to the meta-regression model. In addition to these limitations, the common challenges encountered in prognostic modelling apply. Some prognostic factors may not be available for some individuals or even in whole studies. In this case, multiple imputation methods may be used to improve precision.[55] Finally, numerous candidate prognostic factors might render the available sample size insufficient and model selection challenging.[15]

Further work is needed to extending the model and enhance its flexibility. We developed the baseline risk (Stage 1) in a frequentist framework to take advantage of the software's shrinkage options. However, this might render the results from stage two to be over-precise because the approach does not account for the uncertainty in the baseline risk prediction. An alternative approach would be to carry out a simultaneous estimation of both stages within a Bayesian paradigm. This would allow uncertainty in the estimation during the first stage to be propagated through the model and reflected in the second stage results. Finally, more work is needed to validate both stages of the model. We validated the risk score (Stage 1) only internally, using the bootstrap validation method, but an internal-external validation could also be an option. The predictive accuracy of our two-step framework has not been validated at all. In future work, its performance needs to be validated not only by discrimination and calibration but also metrics related to the absolute benefit.[56]

The proposed approach offers many methodological advantages and opportunities for further development. Model selection approaches and methods to shrink coefficients to avoid extreme predictions are not well established in the meta-analysis context.[1,2] Our proposal shifts the variable selection problem in the logistic regression model for which penalisation methods both in Bayesian and frequentist framework are well established. NMR models can also include aggregated data from published studies, so our approach can be extended accordingly.[43] Observational data can also be integrated to develop the risk score, calibrate or update the risk score model, and externally validate the model. Such data may also inform the baseline effects, or the relative treatment effects and their interactions with the score using appropriate bias-adjusted modelling.[57,58] Methods to include single-arm trials and expert opinion are also available and could be incorporated to extend the model further.[59,60,61,62]

Overall, our framework is flexible enough and combines useful features of predictive modelling and evidence synthesis. It can be applied to as many treatments as required and can be easily extended to include various outcomes. It can inform patients and their doctors, manufacturers, and HTA agencies about the most appropriate treatment for each patient or patients' subgroup and hence contribute to personalised medicine.

**Acknowledgements:** KC, AM and GS are funded by the European Union's Horizon 2020 research and innovation program under grant agreement No 825162. ME was supported by special project funding (grant 189498) from the Swiss National Science Foundation. The authors thank the three anonymous reviewers whose comments/suggestions helped improve and clarify this manuscript. The authors thank Arman Altincatal, Justin Bohn, Shirley Liao, Carl de Moor and Kuangnan Xiong for their comments and their assistance on this article.

**Conflicts of interest:** KC, ES, ME and AM declare that they have no conflict of interest with respect to this paper. FP is an employee of and holds stocks in Biogen; GS was invited to participate in a meeting about real-world evidence organized by Biogen in 2018.

**Data availability statement:** The data that support the findings of this study are available from Biogen International GmbH. Restrictions apply to the availability of these data, which were used under license for this study.

# 5  References

1. Kent DM, Steyerberg E, van Klaveren D. Personalized evidence based medicine: predictive approaches to heterogeneous treatment effects. BMJ. 2018;363:k4245.

2. Rekkas A, Paulus JK, Raman G, et al. Predictive approaches to heterogeneous treatment effects: a scoping review. BMC Med Res Methodol. 2020;20(1):264.

3. Seo M, White IR, Furukawa TA, et al. Comparing methods for estimating patient-specific treatment effects in individual patient data meta-analysis. *Stat Med*. 2021;40(6):1553-1573.

4. Basu A, Carlson JJ, Veenstra DL. A Framework for Prioritizing Research Investments in Precision Medicine. Med Decis Making. 2016;36(5):567-580.

5. Venema E, Mulder MJHL, Roozenbeek B, et al. Selection of patients for intra-arterial treatment for acute ischaemic stroke: development and validation of a clinical decision tool in two randomised trials. The BMJ. 2017;357.

6. Kent DM, Selker HP, Ruthazer R, Bluhmki E, Hacke W. The stroke-thrombolytic predictive instrument: a predictive instrument for intravenous thrombolysis in acute ischemic stroke. Stroke. 2006;37(12):2957-62.

7. Burke JF, Hayward RA, Nelson JP, Kent DM. Using internally developed risk models to assess heterogeneity in treatment effects in clinical trials. Circ Cardiovasc Qual Outcomes. 2014;7(1):163-169.

8. Kent DM, Nelson J, Dahabreh IJ, Rothwell PM, Altman DG, Hayward RA. Risk and treatment effect heterogeneity: re-analysis of individual participant data from 32 large clinical trials. Int J Epidemiol. 2016;45(6):2075-2088.

9. Hayward RA, Kent DM, Vijan S, Hofer TP. Multivariable risk prediction can greatly enhance the statistical power of clinical trial subgroup analysis. BMC Med Res Methodol. 2006;6(1):18.

10. Glasziou PP, Irwig LM. An evidence based approach to individualising treatment. BMJ. 1995;311(7016):1356-1359.

11. Kent DM, Rothwell PM, Ioannidis JP, Altman DG, Hayward RA. Assessing and reporting heterogeneity in treatment effects in clinical trials: a proposal. *Trials*. 2010;11:85. doi:10.1186/1745-6215-11-85

12. Kozminski MA, Wei JT, Nelson J, Kent DM. Baseline characteristics predict risk of progression and response to combined medical therapy for benign prostatic hyperplasia (BPH). BJU Int. 2015;115(2):308-316.

13. Sussman JB, Kent DM, Nelson JP, Hayward RA. Improving diabetes prevention with benefit based tailored treatment: risk based reanalysis of Diabetes Prevention Program. BMJ. 2015 Feb 19;350:h454.

14. Harrell FE. Regression Modelling Strategies: With Applications to Linear Models, Logistic Regression, and Survival Analysis. Springer; 2015.
20

15. Steyerberg EW. Clinical Prediction Models: A Practical Approach to Development, Validation, and Updating. Springer Science & Business Media; 2008.

16. Steyerberg EW, Moons KGM, van der Windt DA, et al. Prognosis Research Strategy (PROGRESS) 3: Prognostic Model Research. PLOS Med. 2013;10(2):e1001381.

17. Moons KGM, Altman DG, Vergouwe Y, Royston P. Prognosis and prognostic research: application and impact of prognostic models in clinical practice. BMJ. 2009;338:b606.

18. Wynants L, van Smeden M, McLernon DJ, et al. Three myths about risk thresholds for prediction models. BMC Med. 2019;17(1):192.

19. Ghasemi N, Razavi S, Nikzad E. Multiple Sclerosis: Pathogenesis, Symptoms, Diagnoses and Cell-Based Therapy. Cell J Yakhteh. 2017;19(1):1-10.

20. Goldenberg MM. Multiple Sclerosis Review. Pharm Ther. 2012;37(3):175-184.

21. Tramacere I DGC, Salanti G DR, Filippini G. Immunomodulators and immunosuppressants for relapsing-remitting multiple sclerosis: a network meta-analysis. Cochrane Database Syst Rev 2015. CD011381(9).

22. McCool R, Wilson K, Arber M, et al. Systematic review and network meta-analysis comparing ocrelizumab with other treatments for relapsing multiple sclerosis. Mult Scler Relat Disord. 2019;29:55-61.

23. Salanti G. Indirect and mixed-treatment comparison, network, or multiple-treatments meta-analysis: many names, many benefits, many concerns for the next generation evidence synthesis tool. Res Synth Methods. 2012;3(2):80-97.

24. Debray TP, Schuit E, Efthimiou O, et al. An overview of methods for network meta-analysis using individual participant data: when do benefits arise? Stat Methods Med Res. Stat Methods Med Res. 2018 May;27(5):1351-1364.

25. Belias M, Rovers MM, Reitsma JB, Debray TPA, IntHout J. Statistical approaches to identify subgroups in meta-analysis of individual participant data: a simulation study. BMC Med Res Methodol. 2019; 19(1):183.

26. Steyerberg EW, Eijkemans MJC, Harrell FE, Habbema JDF. Prognostic modelling with logistic regression analysis: a comparison of selection and estimation methods in small data sets. Stat Med. 2000;19(8):1059-1079.

27. Royston P, Sauerbrei W. Multivariable Model - Building: A Pragmatic Approach to Regression Anaylsis based on Fractional Polynomials for Modelling Continuous Variables. Wiley. 2008.

28. Polman CH, O'Connor PW, Havrdova E, et al. A randomized, placebo-controlled trial of natalizumab for relapsing multiple sclerosis. N Engl J Med. 2006;354(9):899-910.

29. Gold R, Kappos L, Arnold DL, et al. Placebo-controlled phase 3 study of oral BG-12 for relapsing multiple sclerosis. N Engl J Med. 2012;367(12):1098-1107.

30. Fox RJ, Miller DH, Phillips JT, et al. Placebo-controlled phase 3 study of oral BG-12 or glatiramer in multiple sclerosis. N Engl J Med. 2012;367(12):1087-1097.

31. Abadie A, Chingos M M, West MR. Endogenous stratification in randomized experiments. Review of Economics and Statistics. 2018;100(4):567–580.

32. Kent DM, Paulus JK, van Klaveren D, et al. The Predictive Approaches to Treatment effect Heterogeneity (PATH) Statement. Ann Intern Med. 2020;172(1):35-45. doi:10.7326/M18-3667

33. van Klaveren D, Balan TA, Steyerberg EW, Kent DM. Models with interactions overestimated heterogeneity of treatment effects and were prone to treatment mistargeting. J Clin Epidemiol. 2019;114:72-83.

34. Steyerberg EW, Eijkemans MJ, Harrell FE Jr, Habbema JD. Prognostic modeling with logistic regression analysis: in search of a sensible strategy in small data sets. Med Decis Making. 2001;21(1):45-56.

35. Moons KG, Altman DG, Reitsma JB, Ioannidis JP, Macaskill P, Steyerberg EW, Vickers AJ, Ransohoff DF, Collins GS. Transparent Reporting of a multivariable prediction model for Individual Prognosis or Diagnosis (TRIPOD): explanation and elaboration. Ann Intern Med. 2015 Jan 6;162(1):W1-73.

36. Moons KGM, Altman DG, Reitsma JB, et al. Transparent Reporting of a multivariable prediction model for Individual Prognosis or Diagnosis (TRIPOD): explanation and elaboration. Ann Intern Med. 2015;162(1):W1-73.

37. Riley RD, Snell KI, Ensor J, Burke DL, Harrell FE Jr, Moons KG, Collins GS. Minimum sample size for developing a multivariable prediction model: PART II - binary and time-to-event outcomes. Stat Med. 2019;38(7):1276-1296.

38. Kim SM, Kim Y, Jeong K, Jeong H, Kim J. Logistic LASSO regression for the diagnosis of breast cancer using clinical demographic data and the BI-RADS lexicon for ultrasonography. Ultrasonography. 2018;37(1):36-42.

39. Pellegrini F, Copetti M, Bovis F, Cheng D, Hyde R, de Moor C, Kieseier BC, Sormani MP. A proof-of-concept application of a novel scoring approach for personalized medicine in multiple sclerosis. Mult Scler. 2020;26(9):1064-1073.

40. Moons KGM, Donders ART, Steyerberg EW, Harrell FE. Penalized maximum likelihood estimation to directly adjust diagnostic and prognostic prediction models for overoptimism: a clinical example. J Clin Epidemiol. 2004;57(12):1262-1270.

41. Steyerberg EW, Harrell FE, Borsboom GJ, Eijkemans MJ, Vergouwe Y, Habbema JD. Internal validation of predictive models: efficiency of some procedures for logistic regression analysis. J Clin Epidemiol. 2001;54(8):774-781.

42. Steyerberg EW, Harrell FE. Prediction models need appropriate internal, internal-external, and external validation. J Clin Epidemiol. 2016;69:245-247.

43. Saramago P, Sutton AJ, Cooper NJ, Manca A. Mixed treatment comparisons using aggregate and individual participant level data. Stat Med. 2012;31(28):3516-36

44. Donegan S, Dias S, Tudur-Smith C, Marinho V, Welton NJ. Graphs of study contributions and covariate distributions for network meta-regression. Res Synth Methods. 2018;9(2):243-260.

45. Dias S, Ades AE, Welton N, Jansen J, Sutton A. Meta-Regression for Relative Treatment Effects. In: Network Meta-Analysis for Decision Making. John Wiley & Sons, Ltd; 2018:227-271.

46. R Core Team. A language and environment for statistical computing. R Foundation for Statistical Computing, Vienna, Austria. URL https://www.R-project.org/. Published online 2019.

47. Plummer M. JAGS: A Program for Analysis of Bayesian Graphical Models using Gibbs Sampling. 3rd Int Workshop Distrib Stat Comput DSC 2003 Vienna Austria. 2003;124.

48. Yu-Sung Su, Masanao Yajima. R2jags: Using R to Run "JAGS". https://CRAN.R-project.org/package=R2jags. Published online August 23, 2015.

49. Rafiee Zadeh A, Askari M, Azadani NN, et al. Mechanism and adverse effects of multiple sclerosis drugs: a review article. Part 1. Int J Physiol Pathophysiol Pharmacol. 2019;11(4):95-104.

50. Hoepner R, Faissner S, Salmen A, Gold R, Chan A. Efficacy and Side Effects of Natalizumab Therapy in Patients with Multiple Sclerosis. J Cent Nerv Syst Dis. 2014;6:41-49.

51. Lublin FD. Relapses do not matter in relation to long-term disability: no (they do). Mult Scler Houndmills Basingstoke Engl. 2011;17(12):1415-1416.

52. Carl van Walraven. Individual patient meta-analysis--rewards and challenges. *J Clin Epidemiol*. 2010;63(3):235-237. doi:10.1016/j.jclinepi.2009.04.001

53. Sud S, Douketis J. The devil is in the details...or not? A primer on individual patient data meta-analysis. Evid Based Med. 2009;14(4):100-101. doi:10.1136/ebm.14.4.100

54. Stewart LA, Tierney JF. To IPD or not to IPD? Advantages and disadvantages of systematic reviews using individual patient data. Eval Health Prof. 2002;25(1):76-97.

55. Quartagno M, Carpenter JR. Multiple imputation for IPD meta-analysis: allowing for heterogeneity and studies with missing covariates. Stat Med. 2016;35(17):2938-2954.

56. Vickers AJ, Calster BV, Steyerberg EW. Net benefit approaches to the evaluation of prediction models, molecular markers, and diagnostic tests. BMJ. 2016;352:i6.

57. Schnell-Inderst P, Iglesias CP, Arvandi M, et al. A bias-adjusted evidence synthesis of RCT and observational data: the case of total hip replacement. Health Econ. 2017;26 Suppl 1:46-69.

58. Kaiser P, Arnold AM, Benkeser D, et al. Comparing methods to address bias in observational data: statin use and cardiovascular events in a US cohort. Int J Epidemiol. 2018;47(1):246-254.

59. Leahy J, Thom H, Jansen JP, Gray E, O'Leary A, White A, Walsh C. Incorporating single-arm evidence into a network meta-analysis using aggregate level matching: Assessing the impact. Stat Med. 2019;38(14):2505-2523.

60. Signorovitch J, Erder MH, Xie J, Sikirica V, Lu M, Hodgkins PS, Wu EQ. Comparative effectiveness research using matching-adjusted indirect comparison: an application to treatment with guanfacine extended release or atomoxetine in children with attention-deficit/hyperactivity disorder and comorbid oppositional defiant disorder. Pharmacoepidemiol Drug Saf. 2012;21 Suppl 2:130-7.

61. Signorovitch JE, Sikirica V, Erder MH, et al. Matching-adjusted indirect comparisons: a new tool for timely comparative effectiveness research. Value Health J Int Soc Pharmacoeconomics Outcomes Res. 2012;15(6):940-947.

62. Stojadinovic A, Nissan A, Eberhardt J, Chua TC, Pelz JOW, Esquivel J. Development of a Bayesian Belief Network Model for Personalized Prognostic Risk Assessment in Colon Carcinomatosis. Am Surg. 2011;77(2):221-230.

*Appendix*

**Appendix figure 1 Flow-chart for the number of candidate prognostic factors**

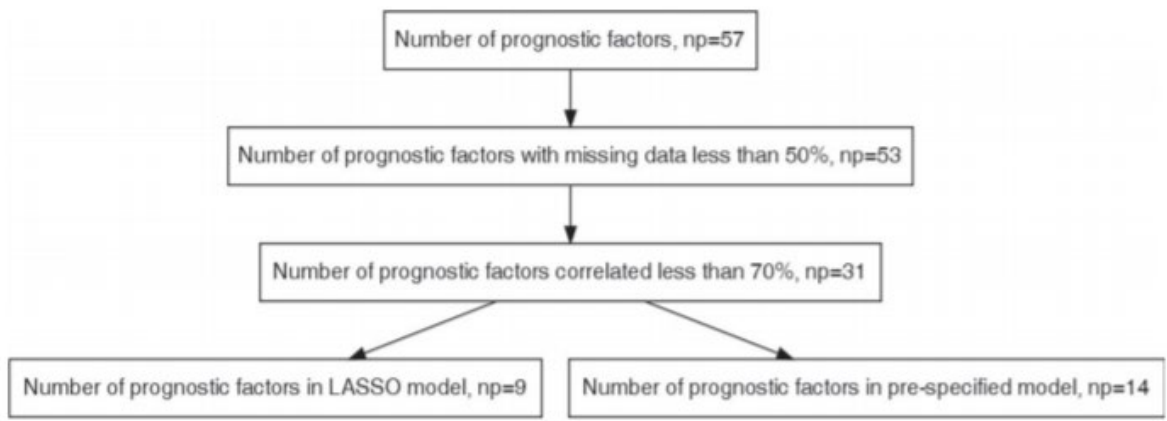

**Appendix figure 2 Venn diagram for candidate characteristics to include in the prognostic model (stage 1). Light blue indicates all 31 characteristics after deleting the correlated variables and those with a big amount of missing values (>50%). Light green indicates the variables selected by LASSO and purple indicates the variables included in pre-specified model**

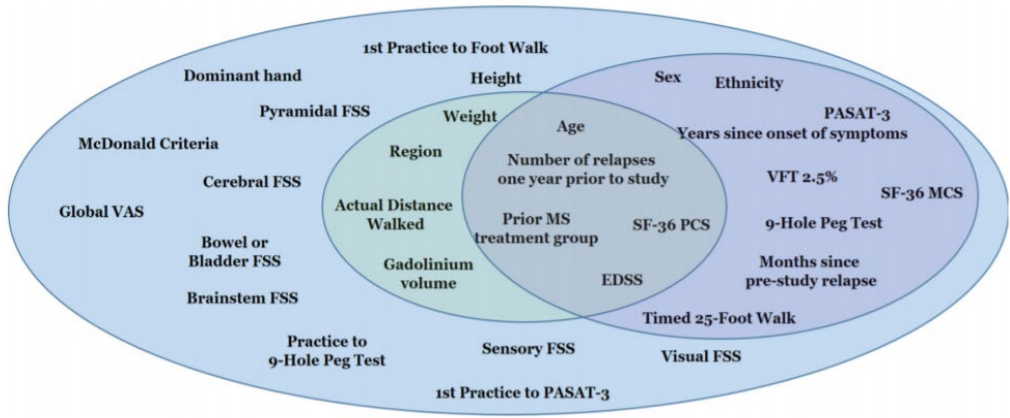

FSS: Functional System Score; VAS: Visual Analog Scale; EDSS: Expanded Disability Status Scale; SF-36 PCS: Short Form-36 Physical Component Summary; SF-36 MCS: Short Form-36 Mental Component Summary; VFT: Visual Function Test.

**Appendix figure 3 ORs of relapse in two years as a function of the baseline risk estimated with LASSO (A) or pre-specified model (B). The x-axis shows the baseline risk score of relapsing in two years. Between the two dashed vertical lines are the baseline risk values observed in our data**

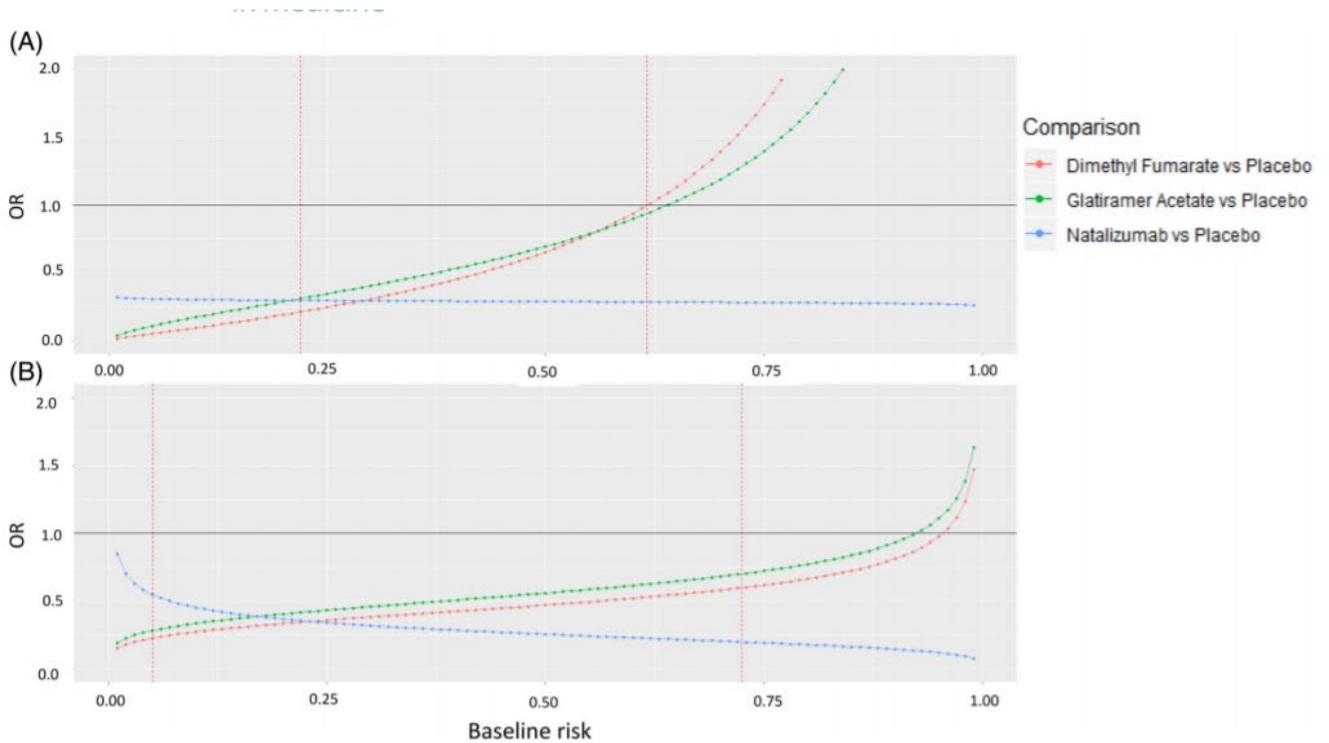